\documentclass{emulateapj}

\usepackage{graphics}
\usepackage{epsfig}
\usepackage{natbib}
\shorttitle{CO Absorption in Protoplanetary Disks}
\shortauthors{McJunkin et al.}

\begin{document}

\title{Probing the Inner Regions of Protoplanetary Disks with CO Absorption Line Spectroscopy}

\author{
Matthew McJunkin\altaffilmark{1}, 
Kevin France\altaffilmark{1}, 
Eric B. Burgh\altaffilmark{1}, 
Gregory J. Herczeg\altaffilmark{2}, 
Eric Schindhelm\altaffilmark{3}, 
Joanna M. Brown\altaffilmark{4}, 
Alexander Brown\altaffilmark{1}
}

\altaffiltext{1}{Center for Astrophysics and Space Astronomy, University of Colorado, 389 UCB,
                         Boulder, CO 80309, USA; matthew.mcjunkin@colorado.edu}
\altaffiltext{2}{Kavli Institute for Astronomy and Astrophysics, Peking University, Beijing 100871, China }                         
\altaffiltext{3}{Southwest Research Institute, 1050 Walnut Street, Suite 300, Boulder, CO 80302, USA} 
\altaffiltext{4}{Harvard-Smithsonian Center for Astrophysics, 60 Garden Street, Cambridge, MA 02138, USA} 
  

\begin{abstract}
	Carbon monoxide (CO) is the most commonly used tracer of molecular gas in the inner regions of protoplanetary disks.  CO can be used to constrain the excitation and structure of the circumstellar environment.  Absorption line spectroscopy provides an accurate assessment of a single line-of-sight through the protoplanetary disk system, giving more straightforward estimates of column densities and temperatures than CO and molecular hydrogen (H$_{2}$) emission line studies.  We analyze new observations of ultraviolet CO absorption from the \emph{Hubble Space Telescope} along the sightlines to six classical T Tauri stars.  Gas velocities consistent with the stellar velocities, combined with the moderate-to-high disk inclinations, argue against the absorbing CO gas originating in a fast-moving disk wind.  We conclude that the far-ultraviolet observations provide a direct measure of the disk atmosphere or possibly a slow disk wind.  The CO absorption lines are reproduced by model spectra with column densities in the range $N$($^{12}$CO) $\sim$ $10^{16}-10^{18}$ cm$^{-2}$ and $N$($^{13}$CO) $\sim$ $10^{15}-10^{17}$ cm$^{-2}$, rotational temperatures $T_{rot}(CO)$ $\sim$ 300~--~700 K, and Doppler $b$-values, $b$ $\sim$ 0.5~--~1.5 km s$^{-1}$.  We use these results to constrain the line-of-sight density of the warm molecular gas ($n_{CO} \sim 70 - 4000$ cm$^{-3}$) and put these observations in context with protoplanetary disk models.    
\end{abstract}

\clearpage

\section{Introduction}

Characterizing the gas properties of protoplanetary disks at planet-forming radii ($a \leq 10$ AU) is critical to understanding the way planets form and evolve.  The lifetimes of protoplanetary disks are comparable to the time during which giant planet cores form and accrete gaseous envelopes ($10^{6}-10^{7}$ yr; \citealt{2010A&amp;A...510A..72F, 2007ApJ...671.1784H, 2005Icar..179..415H}).  The surface density of the disk in which the planet forms places limits on the amount of migration that is possible \citep{2002A&amp;A...394..241T, 2003MNRAS.342.1139A}.  The lifetime of the protoplanetary disk determines the amount of time that giant planets are able to accrete, thereby influencing their final masses \citep{2004ApJ...604..388I}.  Observations of carbon monoxide (CO) emission and absorption lines from ultraviolet (UV) to infrared (IR) wavelengths have been shown to be useful probes of molecular gas in the inner disk \citep{2003ApJ...589..931N, 2011ApJ...743..112S, 2012ApJ...746...97S}.	
	
Emission and absorption from the CO Fourth Positive band system ($A~^{1}\Pi-X~^{1}\Sigma^{+}$) has been widely used  to study the interstellar medium \citep{1980ApJ...242..545F, 2007ApJ...658..446B}, debris disks \citep{1994A&amp;A...290..245V}, comets \citep{1976ApJ...209L..45F, 1997DPS....29.3724M}, planets \citep{1981JGR....86.9115D, 2000ApJ...538..395F}, and the atmospheres of cool stars \citep{1994ApJ...428..329C}.  The first detections of CO far-UV emission and absorption spectral features from the inner regions of protoplanetary disks were presented in \citet{2011ApJ...734...31F}.  The detected CO absorption lines have rotation temperatures $T_{rot}(CO) \approx 500 \pm 200$~K, pointing to an origin in the warm inner disk gas.  Molecular hydrogen (H$_{2}$) and CO absorption were measured simultaneously for the first time by \citet{2012ApJ...744...22F} in the disk of AA Tauri.  The H$_{2}$ was seen in absorption against the Ly$\alpha$ emission line, as in \citet{2011ApJ...730L..10Y}, and the CO $A - X$ absorption bands were observed against the far-UV continuum \citep{2011ApJ...729....7F}.  The value of CO/H$_{2}$ $\sim$~0.4 found by \citet{2012ApJ...744...22F} is approximately three orders of magnitude larger than the canonical interstellar value of $10^{-4}$.  Better constraints on the CO/H$_{2}$ ratio are important for determining the total amount of gas in protoplanetary disks.     
		 	
Many classical T Tauri star (CTTS) disks have been studied with near-IR fundamental (4.7 $\mu$m) and overtone (2.3 $\mu$m) bands of CO, in both emission  (e.g. \citealt{1993ApJ...411L..37C, 2003ApJ...589..931N, 2011ApJ...743..112S}), and absorption (e.g. \citealt{2006ApJ...646..342R, 2012ApJ...754...64H}).  Near-IR CO emission has been used as a tracer of the dominant molecular gas component, H$_{2}$, because the homonuclear nature of H$_{2}$ makes rovibrational transitions dipole-forbidden and thus difficult to observe at near- and mid-IR wavelengths (\citealt{2006ApJ...651.1177P, 2008ApJ...688.1326B, 2008A&amp;A...477..839C}; although see \citealt{2003ApJ...586.1136B, 2007MNRAS.379.1658R}; and \citealt{2008ApJ...678.1088B}). The CO lines at 4.7 $\mu$m appear to originate from the inner disk region with T $\approx $ 500 - 1500 K and Keplerian velocities consistent with the dust sublimation radius. However, some sources have line profiles with excess low velocity emission \citep{2011A&amp;A...527A.119B}, which \citet{2011ApJ...733...84P} propose arises from a slow molecular disk wind.  In general, the IR CO temperatures from LTE slab models (275 - 1675 K, average $\sim 1100$ K; \citealt{2011ApJ...743..112S}) are higher than those measured from  UV CO fluorescence ($460 \pm 250$ K; \citealt{2012ApJ...746...97S}).  \citet{2006ApJ...646..342R} and \citet{2012ApJ...754...64H} studied IR CO absorption from the surfaces of circumstellar disks.  \citet{2006ApJ...646..342R} find CO excitation temperatures of $\sim 100$ K, likely placing the absorbing gas farther out in the disk at $r > 10$ AU.  
     			
The discrepancies between the IR and UV CO work suggests that the observations are probing multiple molecular gas populations, and multiple populations may be responsible for the anomalously high observed CO/H$_{2}$ ratio in AA Tau. With deep UV absorption spectroscopy, we have a new observational tool for studies of the inner disk gas. Absorption lines give the most direct measure of the column density and temperature of the disk gas along the line-of-sight as the line fitting is largely independent of the geometry or the photo-exciting source.  In order to provide observational constraints on inner disk gas, we present new analyses of far-UV spectroscopic observations of six (0.6~--~6 Myr) CTTSs.  Our observations probe the warm disk atmosphere, which is important for constraining the three-dimensional structure of disks in which planets form.  We describe the targets and the observations in $\S 2$.  The analysis performed and the model fit parameters are presented in $\S 3$.  The disk geometry and correlations in the data are discussed in $\S 4$.  Finally, $\S 5$ contains a brief summary of our results.  

\section{Targets and Observations}

	We analyze far-ultraviolet spectra of six targets that are a subset of the 34 T Tauri stars presented in \citet{2012ApJ...756..171F}:  AA Tau, HN Tau A, DE Tau, RECX-15 (ET Cha), RW Aur A, and SU Aur.  HN Tau A and RW Aur A (hereafter HN Tau and RW Aur) are binary stars, but with separations large enough, 1.4" for RW Aur AB and 3.1" for HN Tau AB \citep{2001ApJ...556..265W}, such that only the primary is within the aperture and dominates the emission analyzed here.  AA Tau and HN Tau have been studied previously by France et al. (2011a, 2012a), but we refit their spectra with the same procedure as the other targets for consistency.  Five targets are in the Taurus-Auriga star-forming region at a distance of 140 pc \citep{1999A&amp;A...352..574B, 2007ApJ...671..546L}.  RECX-15 is in the $\eta$ Chamaeleontis cluster at a distance of 97 pc \citep{1999PASA...16..257M}.  Parameters including the age, luminosity, and spectral type of these targets are listed in Table 1. 

\begin{deluxetable*}{ccccccccc}\tablenum{1}
\tabletypesize{\normalsize}
\tablecaption{Target Parameters}
\tablewidth{0pt}
\tablehead{
\colhead{Object }& \colhead{Spectral} & \colhead{A$_{V}$} & \colhead{Inclination} & \colhead{L$_{*}$}& \colhead{M$_{*}$} & \colhead{\.M} & \colhead{log$_{10}$(Age)}&\colhead{Ref.\tablenotemark{a}}\\
&{Type}& &{(degrees)}&{(L$_{\odot}$)}& {(M$_{\odot}$)}&{($10^{-8}$~M$_{\odot}$ yr$^{-1}$)}& {(yrs)} & }

\startdata
AA Tau & K7 & 0.5 & 70 & 0.71 & 0.80 & 0.33 & 6.38~$\pm$~0.20 & 2,4,7,12,16\\
DE Tau & M0 & 0.6 & 35 & 0.87 & 0.59 & 2.64& 5.82~$\pm$~0.20 & 7,10,12\\
HN Tau & K5 & 0.5 & $>$40 & 0.19 & 0.85 & 0.13& 6.27~$\pm$~0.27 & 6,7,12\\
RECX-15 & M2 & 0.0 & 60 & 0.08 & 0.40 & 0.10& 6.78~$\pm$~0.08 &13,14,15\\
RW Aur & K4 & 1.6 & 77 & 2.3 & 1.40 & 3.16& 5.85~$\pm$~0.53 & 5,9,11,12,17\\
SU Aur & G1 & 0.9 & 62 & 9.6 & 2.30 & 0.45& 6.39~$\pm$~0.21 & 1,3,8,11,12\\

\enddata
\tablenotetext{a}{~ (1) \citet{2002ApJ...566.1124A}; (2) \citet{2007ApJ...659..705A}; (3) \citet{1988ApJ...330..350B}; (4) \citet{2010MNRAS.409.1347D}; (5) \citet{2007ApJ...669.1072E}; (6) \citet{2011ApJ...734...31F}; (7) \citet{1998ApJ...492..323G}; (8) \citet{2000ApJ...544..927G}; (9) \citet{1995ApJ...452..736H}; (10) \citet{2001ApJ...561.1060J}; (11) \citet{2000ApJ...539..815J}; (12) \citet{2009ApJ...704..531K}; (13) \citet{2004MNRAS.351L..39L}; (14) \citet{2004ApJ...609..917L}; (15) \citet{2007MNRAS.379.1658R}; (16) \citet{2010A&amp;A...512A..15R}; (17) \citet{2001ApJ...556..265W}.}

\end{deluxetable*}

	All targets were observed using the \emph{Hubble Space Telescope}-Cosmic Origins Spectrograph ($HST$-COS; \citealt{2012ApJ...744...60G}) under program ID 11616.  The COS FUV M-mode wavelength solution is accurate to $\Delta$v $\sim$ 15 km s$^{-1}$ and depends on the object centering\footnote{http://www.stsci.edu/hst/cos/ducuments/handbooks/current/cos\_cover.html}.  Target acquisition was through the MIRRORB near-UV imaging mode for AA Tau and DE Tau.  The rest of the targets were acquired using the MIRRORA mode.  Far-UV spectra were obtained using three central wavelengths for G160M, two central wavelengths for G130M, and multiple focal-plane positions to cover the $1133$ $\leq$ $\lambda$~$\leq$ $1795$~\AA~ bandpass while minimizing fixed pattern noise.  Table 2 lists the  dates and exposure times for each object used in this paper.  The data were processed through CALCOS, the COS calibration pipeline, and aligned and co-added with the procedure described in \citet{2010ApJ...720..976D}.    

\begin{deluxetable*}{cccccc}\tablenum{2}
\tabletypesize{\normalsize}
\tablecaption{$HST$-COS G130M/G160M Observations of Targets}
\tablewidth{0pt}
\tablehead{
\colhead{Object }& \colhead{R. A. (J2000)} & \colhead{Dec. (J2000)} & \colhead{Date} & \colhead{G130M Exposure} & \colhead{G160M Exposure}\\
{}&{}&{}&{}&{(s)}&{(s)} }

\startdata
AA Tau & 04 34 55.42 & +24 28 52.8 & $2011$ Jan $06$, $07$ & $5588$ & $4192$\\
DE Tau & 04 21 55.69 & +27 55 06.1 & $2010$ Aug $20$ & $2088$ & $1851$\\
HN Tau & 04 33 39.37 & +17 51 52.1 & $2010$ Feb $10$ & $5724$ & $4528$\\
RECX-15 & 08 43 18.43 & -79 05 17.7 & $2011$ Feb $05$ & $3890$ & $4501$\\
RW Aur & 05 07 49.51 & +30 24 04.8 & $2011$ Mar $25$ & $1764$ & $1617$\\
SU Aur & 04 55 59.39 & +30 34 01.2 & $2011$ Mar $25$ & $1788$ & $1759$\\
\enddata

\end{deluxetable*}

	The six targets were the only objects in the T Tauri star sample to show unambiguous CO absorption.  The rest of the larger sample either had low continuum signal-to-noise (S/N) making modeling difficult, or were complete non-detections of CO absorption (most likely from low-inclination disks not intercepting the line-of-sight).  Our optical depth detection limit was $\tau = 5$, below which the absorption depths were comparable to the noise in the data.  The targets that we do not fit have inclinations ranging from $4^{\circ}$ (TW Hya, \citealt{2008ApJ...684.1323P}) to $85^{\circ}$ (DF Tau, \citealt{2001ApJ...561.1060J}) with an average of $43^{\circ}$.  Thus, the majority of the targets that do not show strong CO absorption have lower inclinations.  DF Tau is an unusual target discussed in $\S 4.3$.  The six targets that we analyze all have medium-to-high inclination ($35^{\circ}$ - $77^{\circ}$, with an average of  $61^{\circ}$) protoplanetary disks.  The average does not include HN Tau, whose inclination is not well known.  The inclinations have been determined in a number of different ways, with fairly large uncertainties arising from the techniques used.  The inclination of AA Tau was determined by periodic eclipses of the star by the magnetically warped accretion disk combined with a measured line-of-sight projected rotation velocity in the optical \citep{2010MNRAS.409.1347D, 2003A&amp;A...409..169B}.  For DE Tau, it was calculated from stellar radii and rotation periods spectroscopically determined in the IR combined with the literature value of $v$~sin~$i$ in the red \citep{2001ApJ...561.1060J}.  Optical spectroscopy of a very strong H$_{\alpha}$ emission line profile was modeled to find the inclination of the disk of RECX-15 \citep{2004MNRAS.351L..39L}.  The interferometric data of RW Aur was fit with an inclined uniform disk model and was sufficient to constrain the inclination of the disk \citep{2007ApJ...669.1072E}.  This inclination is inconsistent with the RW Aur inclination ($\sim 45^{\circ}$) found by \citet{2003A&amp;A...405L...1L} using the ratio of proper motion to radial velocity toward emission knots in the RW Aur jet.  The visibility as a function of hour angle in the K-band of SU Aur was fit with a Gaussian brightness profile that was inclined on the sky, giving the best fit inclination for the disk \citep{2002ApJ...566.1124A}.  The inclination of HN Tau is not well known, and only a lower limit is given to indicate that an inclination higher than 40$^{\circ}$ is needed for the line-of-sight to intercept the disk and to explain observed outflows \citep{2011ApJ...734...31F}.  This lower limit is increased in $\S 4.3$ by comparing to published models.  
	
	The radial velocities of the majority of the stars (listed in Table 3) were obtained by \citet{2012ApJ...745..119N}.  The radial velocity for RW Aur A was not measured by \citet{2012ApJ...745..119N}, but is assumed to be the same as that for the secondary, RW Aur B. The radial velocity of RECX-15 is determined using photospheric Li I absorption lines at 6707.76~\AA~ and 6707.91~\AA~ \citep{2011A&amp;A...534A..44W} with resolving power of $\sim 30,000$ (10 km s$^{-1}$ at 6300~\AA).                 
	  
	The CO is observed in absorption against the far-UV continuum of these actively accreting stars.  The far-UV continuum arises from stellar photospheric and chromospheric emission as well as an accretion continuum (\citealt{1998ApJ...509..802C, 2011ApJ...729....7F, 2011ApJ...743..105I}; see Table 1 for the accretion rates).  The far-UV continuum may be enhanced in some objects by a molecular dissociation quasi-continuum \citep{2004ApJ...614L.133B, 2004ApJ...607..369H}, but the details of this process have not been thoroughly characterized observationally (see, e.g., \citealt{2011ApJ...729....7F}).  The continuum emission is absorbed by circumstellar atoms and molecules.  Depending on the inclination angle, the light may pass through multiple layers of the disk and thus probe multiple temperature and density regimes.  At the high temperatures and column densities of the observed CO (see $\S 3.3$), the contribution from interstellar CO is negligible for transitions from $J$ $> 2$.  For temperatures typical of the interstellar medium (ISM), states with $J$ $\ge$ 3 are not significantly populated.  For our extinction values (A$_{V} \lesssim 1.5$), we would expect an interstellar CO column density of $N$(CO) $\lesssim 10^{14}$ (see Figure 3 of \citealt{2007ApJ...658..446B}), which is orders of magnitude smaller than the circumstellar column densities that we observe.  Therefore, our CO measurements do not probe the cold diffuse interstellar material with $<$$T_{rot}(CO)$$>$$_{ISM}$ $\sim$ 4 K \citep{2007ApJ...658..446B}, but arise from material close to the star, likely the warm molecular disk.

\section{Analysis}

\subsection{Data Analysis}
	We identify nine CO bands, ($v^{'}$ - 0), that span the COS far-UV
	bandpass from the (0 - 0) band at $\lambda$~$\sim$~$1544$~\AA~ to
	the (8 - 0) band at $\lambda$~$\sim$~$1322$~\AA~ (Figure 1).  We do
	not fit the (5 - 0) band due to contaminating emission from H$_{2}$
	and Si IV $\lambda 1394$~\AA.  The (0 - 0) band is contaminated
	with C IV $\lambda 1548$~\AA~ emission and is also excluded from
	the overall fit.  Many photo-excited H$_{2}$ emission lines
	populate the spectra, complicating the model spectrum fits as well.
	All CTTS systems with identifiable CO absorption contain broad
	bands created by the overlap of many closely spaced rotational
	lines.  This suggests a relatively high rotational temperature,
	$T_{rot}(CO)$, because as higher-$J$ lines are populated, the
	observed spectral bands become broader.  Broad CO absorption bands
	were first identified in the spectrum of HN Tau
	\citep{2011ApJ...734...31F} and later in the spectrum of AA Tau
	\citep{2012ApJ...744...22F}.  We employ a similar technique to the
	previous work, using spectral synthesis modeling to measure
	$T_{rot}(CO)$ and N(CO) in all six sources.  

\begin{figure*} \figurenum{1}
\begin{center}
\epsfig{figure=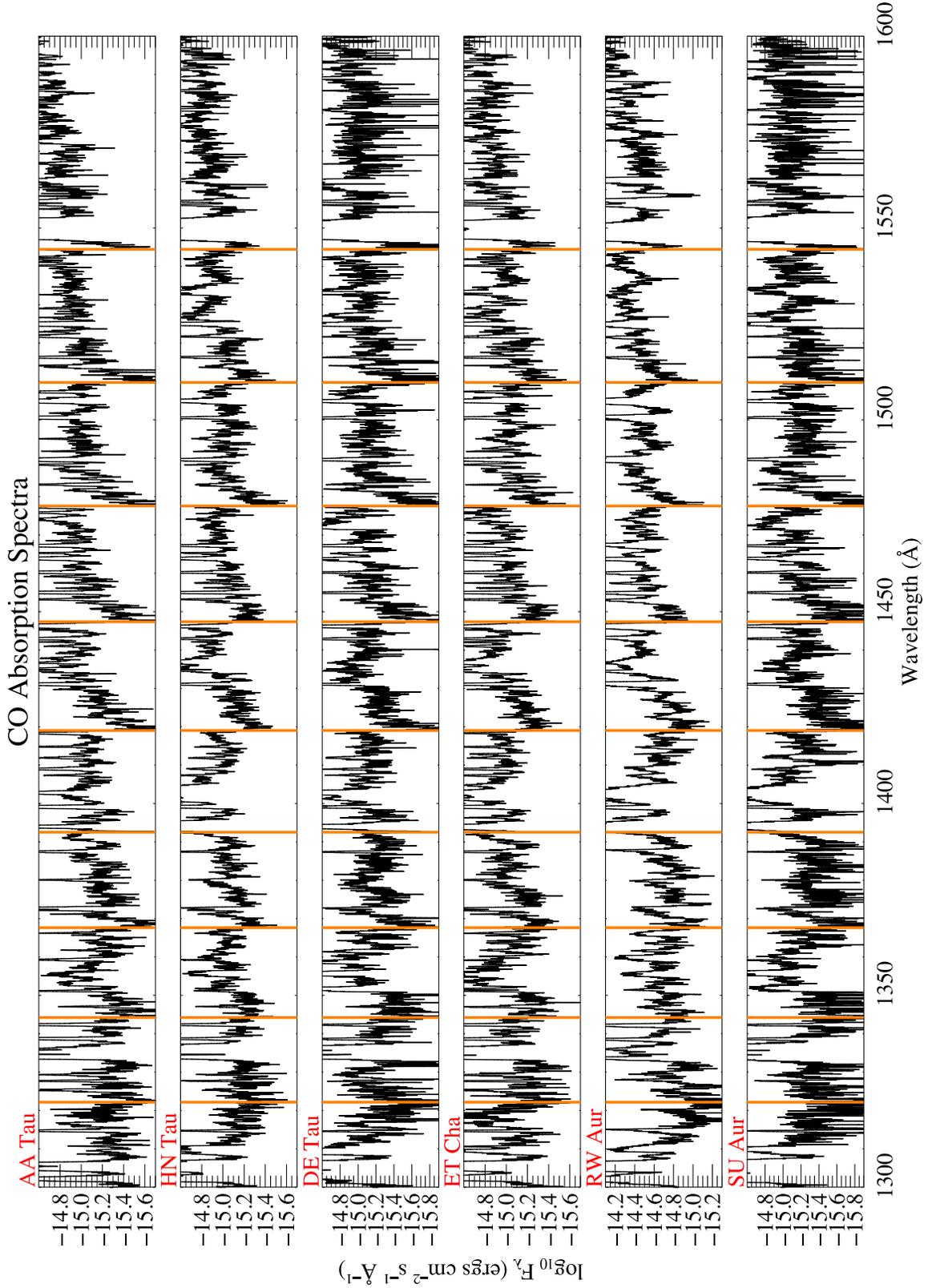,scale=0.9}

\caption{
The nine far-UV CO rovibrational absorption bandheads identified by orange vertical lines.  Going from left to right: ($v^{'}$ - $v^{''}$ = 8 - 0) $1322.1$~\AA, (7 - 0) $1344.2$~\AA, (6 - 0) $1367.6$~\AA, (5 - 0) $1392.5$~\AA, (4 - 0) $1419.0$~\AA, (3 - 0) $1447.4$~\AA, (2 - 0) $1477.6$~\AA, (1 - 0) $1509.7$~\AA, (0 - 0) $1544.4~$\AA.  The spectra have been Fourier smoothed to suppress frequencies higher than the 7-pixel spectral resolution.
}
\end{center}
\end{figure*}

	The spectrum around each band was normalized by a linear fit to the nearby continuum, avoiding regions with H$_{2}$ or other emission lines.  The least contaminated absorption bands are the (1 - 0), (2 - 0), (3 - 0), and (4 - 0) bands, which also have the strongest oscillator strengths in the Fourth Positive system.  However, $^{12}$CO and $^{13}$CO are not cleanly resolved in the low-$v^{'}$ bands.  To isolate these isotopologues, higher-$v^{'}$ bands with larger separations between the $^{12}$CO and $^{13}$CO bandheads ($v^{'}$ = 7, 8) are used to better constrain the column densities of both species (see $\S 3.3$).  There is nearly a 3~\AA~ separation between the isotopes in the (7 - 0) band, compared to the $\sim$1~\AA~ separation in the (1 - 0) band.  However, these higher bands can sometimes blend into the continuum because of their shallower absorptions.  HN Tau in particular does not have detected $v^{'}$ = 7, 8 absorption (see \S 3.3).  An example of these higher band fits showing separations in the DE Tau (1 - 0) and (7 - 0) lines is shown in Figure 2.  

\begin{figure} \figurenum{2}
\begin{center}
\plotone{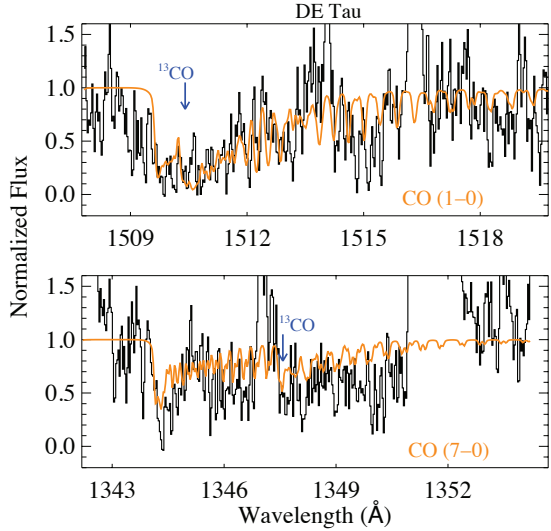}

\caption{
Low-$v^{'}$ (above) and high-$v^{'}$ (below) model fits for DE Tau showing the larger separation between $^{12}$CO and $^{13}$CO for larger $v^{'}$-values.  The separation helps to better constrain the $^{12}$CO/$^{13}$CO ratio in the model.  The data is in black and the best-fit model is in orange.  The first absorption dip in each model is the $^{12}$CO bandhead, and the $^{13}$CO bandhead is marked with a blue arrow. 
}
\end{center}
\end{figure}
	
	The normalized data vary in both the depth of their absorptions and the number of excited $J$-lines present.  DE Tau and SU Aur have the largest depths, but the data are also noisier, making it harder to determine a continuum level.  The depths will vary, even for similar column densities, because of the error associated with the continuum fit and population shifts in the lower-$J$ levels due to increases and decreases in temperature (see \S 3.2).  Because the absorptions in these targets are very saturated (see $\S 3.3$), changing the column density does not affect the absorption depth highly at low-$J$ values.  Here, the normalization will have a bigger effect.  At high-$J$ values (longer wavelengths within the band) it is difficult to pinpoint the weaker absorptions amongst the emission lines and the noise in the data.  This makes the data harder to interpret, though RW Aur does have a significant lack of high-$J$ lines compared to the other targets.  HN Tau, however, has fairly prominent high-$J$ lines.  All of these factors contribute to the derived molecular parameters and errors described in \S 3.2 and \S 3.3.        
	
\subsection{Model Description and Fitting Procedure}
	The CO absorption profiles were modeled for $^{12}$CO using the oscillator strengths and rovibrational line wavelengths from Eidelsberg (private communication).  The A $^{1}\Pi$ energy levels of \citet{1994ApJ...420..433H} were used to calculate the ground state energies.  The oscillator strengths for $^{13}$CO are taken from \citet{1999A&amp;A...346..705E}, and the wavelengths and oscillator strengths for the perturbation states were taken from \citet{2003ApJS..145...89E}.  A common rotational temperature was assumed for both $^{12}$CO and $^{13}$CO. 

	All six of the analyzed bands ($v^{'}$ = 1, 2, 3, 4, 7, 8) were compared to model spectra for a grid of different values of the Doppler $b$-value, rotational temperature, $T_{rot}(CO)$, and logarithmic column density of both $^{12}$CO and $^{13}$CO, log$_{10}$($N$($^{12}$CO)) and log$_{10}$($N$($^{13}$CO)).  The model also includes a velocity shift parameter (explored in more detail in \S 4.1), which we originally estimated by eye from the (2 - 0) band of each target by varying the velocity shift until the bandheads of the model and data coincided.  However, after a more rigorous treatment of the velocity shifts, using all the bands with uncontaminated bandheads to constrain the location of the CO gas in the disk, more reliable values are found in \S 4.1.  The ranges of our grid search were 0.5~--~2.0 km s$^{-1}$ in steps of 0.1 km s$^{-1}$ for the Doppler $b$-value,  300~--~1000 K with steps of 50 K for the rotational temperature, 16.0~--~18.0 in steps of 0.1 for log$_{10}$($N$($^{12}$CO)), and 14.0~--~17.0 in steps of 0.1 for log$_{10}$($N$($^{13}$CO)).  The maximum  $b$-value for our grid search comes from our assumption of turbulent velocities $\lesssim 1$ km s$^{-1}$ and CO rotational temperatures $\lesssim 5\times 10^{3}$ K.  The maximum log$_{10}$($N$($^{12}$CO)) value for the search comes from the analysis of \citet{2012ApJ...744...22F}, who argue that any larger column of $^{12}$CO would lead to a column of hydrogen that would extinguish the stellar flux around the Lyman alpha line center, which is inconsistent with our observations.  This limit is confirmed after the fit as the majority of the targets (except those with the highest column, RECX-15 and SU Aur) are inconsistent with a column of $10^{18}$ cm$^{-2}$ within their errors.  Any $^{12}$CO column density lower than our lower limit of $10^{16}$ cm$^{-2}$ (which corresponds with our optical depth detection limit of $\tau = 5$) would be very difficult to detect with the S/N, setting the lower bound on our column density search.  An example of our CO model is shown in Figure 3.  The higher rotational states are excited at higher temperatures while the lower rotational states are depopulated.  Thus, the observed absorption band becomes wider and shallower with increasing temperature if column density is kept constant.  The three simulated spectra at the top of Figure 3 are at the native model resolution ($\Delta$v $\sim$ 2 km s$^{-1}$) and the bottom spectrum is at the resolution of the $HST$-COS instrument ($\Delta$v $\sim$ 17 km s$^{-1}$).  The CO absorption lines are narrower than the instrumental resolution.  The unresolved line cores do not go to zero at the observational resolution even when the CO optical depths are large.  

\begin{figure} \figurenum{3}
\begin{center}
\plotone{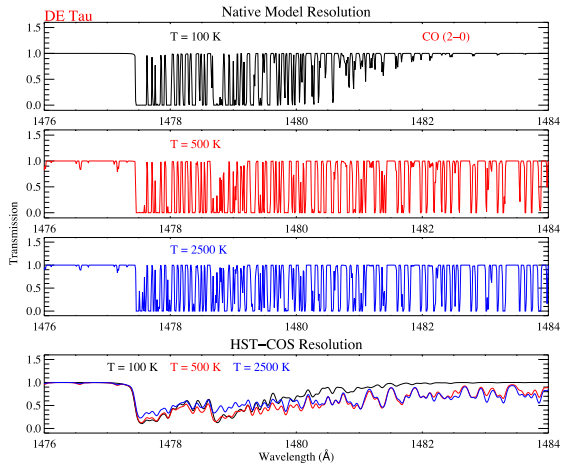}

\caption{
Three different CO models for the (2 - 0) band of DE Tau with identical Doppler $b$-value~$=1.4$ km s$^{-1}$, log$_{10}$($N$($^{12}$CO))~$=17.2$ , log$_{10}$($N$($^{13}$CO))~$=16.0$, and velocity shift~$=1.4$ km s$^{-1}$ .  (Top 3): Model output with native model resolution ($\Delta$v $\sim 2$ km s$^{-1}$).  (Bottom): Model output after being convolved with the $HST$-COS linespread function ($\Delta$v $\sim 17$ km s$^{-1}$) to simulate data detected with the instrument.     
}
\end{center}
\end{figure}
		
	 Because the statistical errors in the data were sometimes anomalously small\footnote{The CALCOS pipeline is known to mishandle statistical errors in the low S/N regime \citep{2011ApJ...743...26F}}, such that a reduced $\chi^{2}$ fitting routine with non-uniform errors gave erroneous results, we employed a simple least squares fitting routine that minimizes the square of the difference between the model and the data.  This procedure weighted all data points equally, giving a more reliable fit.  Because of the presence of emission lines (mostly H$_{2}$), we truncated the data above a normalized flux level of $1.1$ so that the emission features did not contaminate the absorption model fitting.  This does, however, cause the model to fit a truncated continuum value instead of the absorption at times when the emission lines wipe out the absorption.  The model is pulled toward a solution with lower column density to fit the higher flux values at these points, introducing a bias which contributes to the large errors on the column density in Table 3.               

	The model uses the five input parameters to create a theoretical spectrum.  The Doppler $b$-value, along with the column density, determines the opacity of the model, and the temperature determines the band shape as discussed above, with higher temperatures leading to wider, shallower absorption models.  The column density inputs determine the relative depth of the $^{12}$CO and $^{13}$CO absorption bands.  
The $^{12}$CO/$^{13}$CO ratio is primarily determined by the differences in the absorption depths at the relevant bandheads.  At the high column densities of these disks (see $\S 3.3$), most of the targets are on the flat portion of the curve of growth (COG) for the lower-$J$ $^{12}$CO lines of the lower-$v^{'}$ bands.  This means that increasing the column density of $^{12}$CO does not lead to a significant increase in the absorbed flux.  The column density fits rely more on the $^{13}$CO absorption and the higher-$v^{'}$ bands and higher-$J$ lines of $^{12}$CO, which are less saturated.  The velocity shift parameter moves the absorption band of the model to shorter or longer wavelengths corresponding to the gas moving toward or away from the observer, respectively.  The theoretical spectrum is then convolved with the COS linespread function and compared to the observations to find the best-fit parameter values.             

\subsection{Model Fits and Errors}
	The fits to the normalized absorption data are shown in Figure 4
	and the parameter values are provided in Table 3.  DE Tau and SU
	Aur, which have the deepest absorptions, have higher column
	densities in both $^{12}$CO and $^{13}$CO than most of the other
	targets.  The shallower low-$J$ lines combined with the prominent
	high-$J$ lines in HN Tau require a higher temperature and lower
	column density to fit the data.  Most of the targets share a common
	best-fit temperature ($\sim 350$ K), with HN Tau being an outlier
	because of the more prominent high-$J$ lines.  The column densities
	are much more varied, with large uncertainties because of the
	high-$J$ lines blending in with the low S/N continuum and
	contamination by emission lines.  Unlike the temperature, there is
	no common best-fit column density for all of the targets within the
	uncertainties.  An intersystem band not taken into account in the
	code ($a^{'}~^{3}\Sigma^{+}-X~^{1}\Sigma^{+}$ (14 - 0)) at $\lambda
	= 1419.50$~\AA~\citep{2003ApJS..145...89E} hinders the fit of the
	(4 - 0) band of HN Tau, the highest temperature target.  This
	absorption is between the $^{12}$CO and $^{13}$CO bandheads,
	causing the data to be deeper than the model in this part of the
	spectrum.  A few targets also show this difference in the (2 - 0)
	band as well, which could be another intersystem band. 

\begin{deluxetable*}{ccccccccc}\tablenum{3}
\tabletypesize{\footnotesize}
\tablecaption{CO Fit Parameters}
\tablewidth{0pt}
\tablehead{
\colhead{Object} & \colhead{F$_{cont}$(1420)} & \colhead{log$_{10}$($N$($^{12}$CO))} & \colhead{log$_{10}$($N$($^{13}$CO))} &\colhead{$T_{rot}(CO)$} & \colhead{$b$} & \colhead{v$_{CO}$} & \colhead{v$_{*}$} & \colhead{v$_{*}$ Reference\tablenotemark{a}}\\
 &(10$^{-15}$~erg cm$^{-2}$ s$^{-1}$\AA$^{-1}$)& & &(K)&(km s$^{-1}$)&(km s$^{-1}$)&(km s$^{-1}$)& }

\startdata
AA Tau & 6.1 & $16.9^{+0.6}_{-0.7}$ & $15.5^{+1.0}_{-1.2}$ & $450^{+550}_{-250}$ & $1.4^{+0.7}_{-0.6}$ & $+24.7\pm18.6$ & $16.98\pm0.04$ & 2\\
DE Tau & 3.7 & $17.2^{+0.6}_{-0.7}$ & $16.0^{+1.0}_{-1.7}$ & $450^{+1000}_{-300}$ & $1.4^{+0.6}_{-0.5}$ & $
+1.4\pm15.9$ & $15.402\pm0.018$ & 2\\
HN Tau & 7.9 & $16.0^{+0.4}_{-0.6}$ & $15.2^{+0.7}_{-1.2}$ & $700^{+1200}_{-450}$ & $1.4^{+1.3}_{-0.6}$ & $+28.0\pm21.1$ & $4.6\pm0.6$ & 2\\
RECX-15 & 4.5 & $17.8^{+0.6}_{-0.9}$ & $16.7^{+1.2}_{-1.0}$ & $350^{+1850}_{-200}$ & $0.5^{+0.3}_{-0.2}$ & $+18.2\pm11.3$ & $22$ & 1\\
RW Aur & 22.3 & $16.9^{+0.8}_{-1.0}$ & $14.7^{+0.7}_{-0.7}$ & $300^{+600}_{-250}$ & $0.6^{+0.4}_{-0.2}$ & $
+20\pm20$ & $15.00\pm0.03$\tablenotemark{b} & 2\\
SU Aur & 5.6 & $17.4^{+0.7}_{-0.9}$ & $15.5^{+1.4}_{-1.5}$ & $350^{+1250}_{-250}$ & $1.2^{+0.7}_{-0.5}$ & $+16.3\pm27.3$ & $14.26\pm0.05$ & 2\\

\enddata
\tablenotetext{a}{ (1) \citet{2011A&amp;A...534A..44W}; (2) \citet{2012ApJ...745..119N}.  }
\tablenotetext{b}{ Value for RW Aur B }
\end{deluxetable*}

\begin{figure} \figurenum{4a}
\begin{center}
\plotone{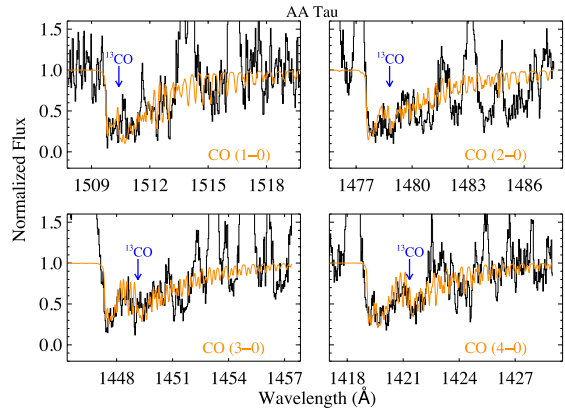}

\caption{
Low-$v^{'}$ model fits (orange) for AA Tau.  The best-fit values (see Table 3) for the Doppler $b$-value, $T_{rot}(CO)$, log$_{10}$($N$($^{12}$CO)) , log$_{10}$($N$($^{13}$CO)), and velocity shift are used.  The data (black) are continuum normalized and the $^{13}$CO bandheads are marked to clearly show the two different CO species.  The emission contaminating the absorption is not included in the model and any flux above a normalized value of 1.1 is truncated during the fit.    
}
\end{center}
\end{figure}

\begin{figure} \figurenum{4b}
\begin{center}
\plotone{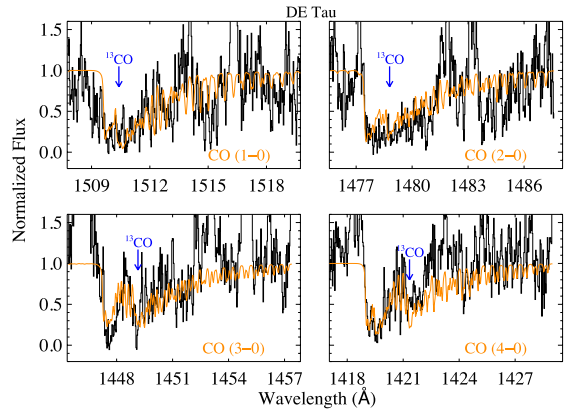}

\caption{
Same as Figure 4A for DE Tau.
}
\end{center}
\end{figure}

\begin{figure} \figurenum{4c}
\begin{center}
\plotone{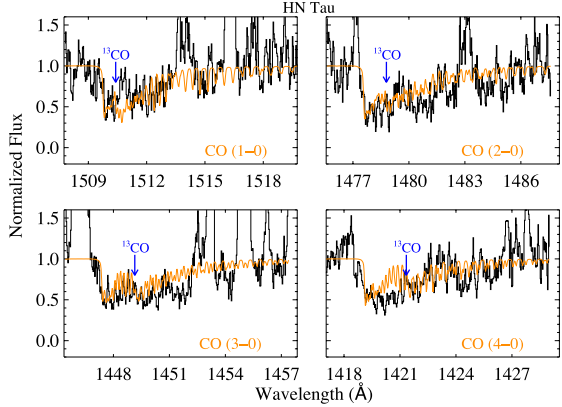}

\caption{
Same as Figure 4A for HN Tau.
}
\end{center}
\end{figure}

\begin{figure} \figurenum{4d}
\begin{center}
\plotone{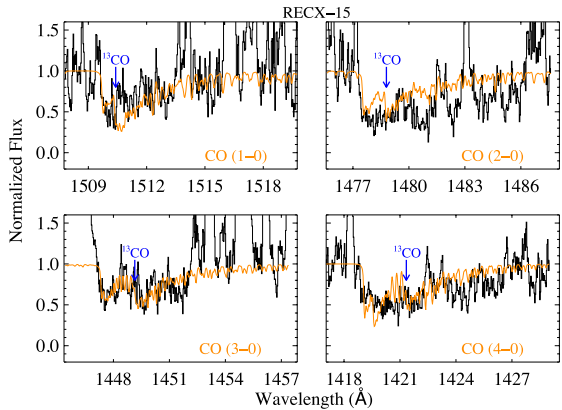}

\caption{
Same as Figure 4A for RECX-15.
}
\end{center}
\end{figure}

\begin{figure} \figurenum{4e}
\begin{center}
\plotone{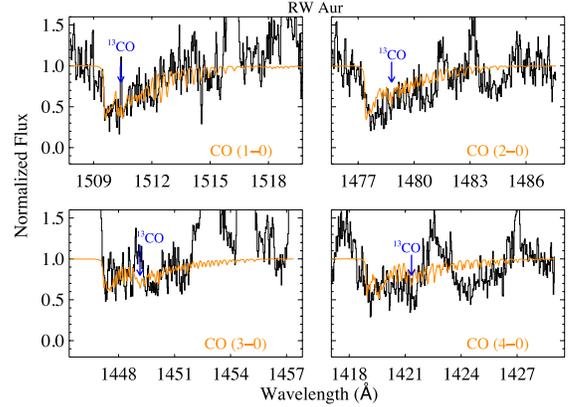}

\caption{
Same as Figure 4A for RW Aur.
}
\end{center}
\end{figure}

\begin{figure} \figurenum{4f}
\begin{center}
\plotone{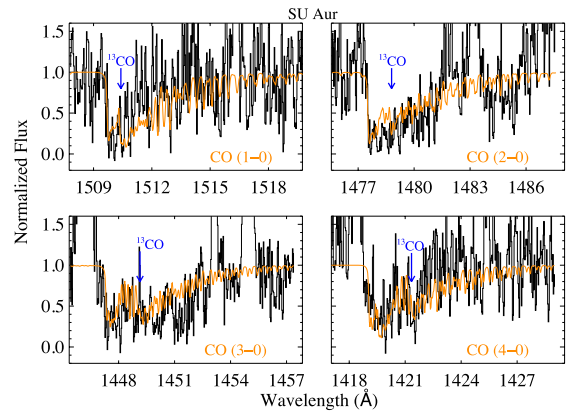}

\caption{
Same as Figure 4A for SU Aur.
}
\end{center}
\end{figure}

	Our fit values for all six targets lie in the ranges of $b$ = 0.5~--~1.4 km s$^{-1}$, $T_{rot}(CO)$ = 300~--~700 K, log$_{10}$($N$($^{12}$CO)) = 16.0~--~17.8, and log$_{10}$($N$($^{13}$CO)) = 14.7~--~16.7.  
The errors were estimated by taking the standard deviation of the $\chi^{2}$ distribution with appropriate degrees of freedom.  Increasing our minimum $\chi^{2}$ value by the standard deviation determined the best-fit parameter range.  We define errors as the width of the best-fit parameter space, as defined by the standard deviation of the $\chi^{2}$ distribution.  This error procedure is illustrated in Figure 5.  The errors on log$_{10}$($N$($^{13}$CO)) were more difficult to define because of the lower column density of $^{13}$CO.  For log$_{10}$($N$($^{13}$CO)) $\lesssim$ 14, the $\chi^{2}$ value does not increase substantially as the column density is decreased because we are no longer detecting the absorption.  We take a lower limit on the column density of $^{13}$CO of $10^{14}$ cm$^{-2}$ for this reason.  Within the errors, HN Tau, RW Aur, and SU Aur are consistent with a non-detection of $^{13}$CO (column density of $^{13}$CO $<$ $10^{14}$ cm$^{-2}$). 
The temperatures in general tend to have large upper error bars.  Higher temperatures populate the higher $J$-states, which produce weaker absorption lines due to smaller column densities.  Due to the low S/N in the data, these weak lines blend into the continuum making the high temperature limit difficult to constrain.  At temperatures $\gtrsim$ 1000 K, the ($v^{'}$ - 1) rovibrational bands of CO should be detectable ($\tau \geq 5$) in the data at our best-fit column densities.  At our best-fit temperature of $\sim 350$ K, the ($v^{'}$ - 1) band would be detected for log$_{10}$($N$($^{12}$CO))  $\gtrsim$ 18.0.  However, we do not detect any of these bands, which suggests that the CO gas is not at densities much greater than $10^{18}$ cm$^{-2}$, or at temperatures much greater than 1000 K, though our models of the ($v^{'}$ - 0) bands alone do not provide strong constraints on the high end of the temperature range.  We find similar results within the errors for AA Tau to the absorption line analysis of \citet{2012ApJ...744...22F}.  Our $b$-values are much higher than those found with sub-millimeter observations of the disk of TW Hya ($b$ $\lesssim$ 0.04 km s$^{-1}$) by \citet{2011ApJ...727...85H}, however, this could be a consequence of their observations probing colder, more quiescent CO gas at disk radii of $\sim 100$ AU.  

\begin{figure} \figurenum{5}
\begin{center}
\plotone{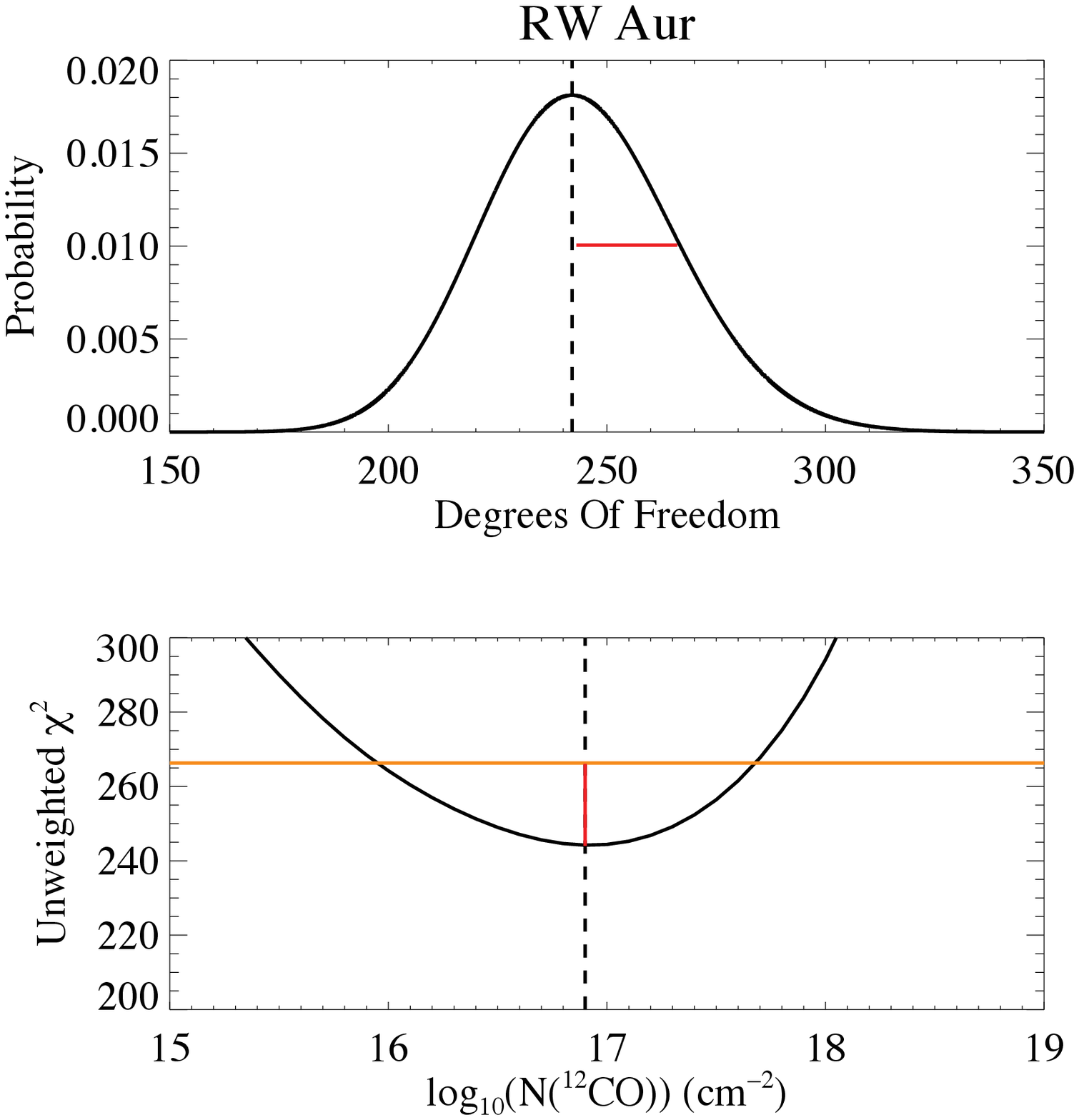}

\caption{
Illustration of the uncertainty determination.  (Above): A plot of the $\chi^{2}$ distribution for the degrees of freedom of RW Aur.  The standard deviation of the distribution is shown in red; the mean is defined by the dashed line.  (Below):  A plot of the unweighted $\chi^{2}$ value as a function of log$_{10}$($N$($^{12}$CO)).  We increased the minimum $\chi^{2}$ by the standard deviation of the distribution to define our error region.  The dashed line shows the best-fit column density for RW Aur ($10^{16.9}$ cm$^{-2}$), the red line shows the standard deviation of the above $\chi^{2}$ distribution, and the orange line defines the error region where it crosses the unweighted $\chi^{2}$ line.   
}
\end{center}
\end{figure}
	
		
	The errors are driven mainly by three sources: continuum determination, low S/N, and saturation in the absorption lines.  Estimates of the continuum fluxes are difficult due to the large number of emission lines in these spectra, especially the emission coincident with the absorption bands.  The error in the determination of the continuum flux is $\sim 20$\%, leading to an error on the column densities of $\sim 0.25$ dex.  The low S/N in the continuum (S/N per resolution element $\sim$ 7 for AA Tau) and absorption lines also contribute to the errors.  Typical continuum flux values at  $\lambda 1420$~\AA~ are listed in Table 3.  At these low fluxes, the absorbers could not be detected with the Space Telescope Imaging Spectrograph (STIS), and have only recently been detected with COS.  Saturation in these lines also leads to a degeneracy between the column densities and the Doppler $b$-values.  We include higher-$v^{'}$ bands in our fit as they have smaller oscillator strengths and are therefore less saturated. 
			
	The line optical depths for all the targets (except HN Tau) are of order 100~--~600 for the low-$v^{'}$ transitions ($v^{'}$ = 0 - 6) up to $J$-values of 15 to 25, which puts these lines on the flat portion of the COG.  HN Tau, however, has optical depths of about 20 or less for all $v^{'}$ transitions and even reaches optical depths of order unity in the models for the (7 - 0) band and therefore was undetected in the observations.  The lower optical depths are caused by the higher temperature and higher $b$-value of HN Tau.  The higher temperature (700 K) of HN Tau lowers the line strength of each $J$-line compared to if the gas was cooler by distributing the absorption over more transitions.  Similarly, the higher $b$-value (1.4 km s$^{-1}$) lowers the line center cross-section through larger Doppler broadening spreading out the cross-section in wavelength space.  The product of the lower line strength and lower cross-section in each $J$-line leads to a significantly lower optical depth for HN Tau.  None of the other targets approach optical depths as small as 1 until the high-$J$ levels of the $v^{'}$ = 7, 8 states.   
	
\section{Results and Discussion}
\subsection{CO Velocity and Isotopic Fraction}
	
	The radial velocities of the CO absorption lines listed in Table 3 were obtained with a simple least squares fit.  Assuming the best-fit parameters of the CO model, velocity shifts from -200 km s$^{-1}$ to +200 km s$^{-1}$ with 1 km s$^{-1}$ intervals were explored.  Due to low S/N, the (8 - 0) band was not used in the velocity calculation.  The average and standard deviation of the velocity shifts for the observed bands were taken as each target's velocity and velocity error.  Shifting the velocity moves the model high-$J$ CO lines relative to the observed spectra, which can change the best-fit temperature.  However, the temperature would change by only $\sim 50$ K, which is well within our errors.  The range of acceptable velocities was small enough that the best-fit parameters from the model were not affected by the shift.       
	
	 Since RW Aur contains strong, redshifted H$_{2}$ emission lines \citep{2012ApJ...756..171F}, which sometimes places the emission on top of the CO absorption bandhead, we checked the velocity shift by hand.  We required that the bandhead of the model and data match for the (1 - 0), (2 - 0), and (4 - 0) bands, which are the cleanest and strongest absorption bands.  The best-fit velocity shift is +20 km s$^{-1}$, consistent with the stellar radial velocity (+15 km s$^{-1}$) in the literature \citep{2012ApJ...745..119N}.  The fit starts to become noticeably worse at $\pm 20$ km s$^{-1}$ from the best-fit value, which we take as our error.   
	 
	 The fitted CO absorption line velocities are generally consistent with the radial velocities of the stars from the literature to within the 15 km s$^{-1}$ absolute uncertainty in the COS wavelength scale.  However, the CO absorption in DE Tau appears to be somewhat blueshifted relative to the stellar velocity (see Table 3), but is still consistent within the errors.  As noted by \citet{2012ApJ...745..119N}, the low stellar radial velocity of HN Tau (4.6 km s$^{-1}$) deviates from the average velocity of the Taurus-Auriga star-forming region ($\sim$ 15 km s$^{-1}$; \citealt{1986ApJ...309..275H}).  The quoted stellar radial velocity of HN Tau is inconsistent with our CO velocity, but is only slightly outside our errors.  These small velocity differences between the CO absorption and the star ($\Delta$v = 2.04 - 23.4 km s$^{-1}$) indicate that the CO is approximately at rest in the stellar frame, or at least not in a fast-moving disk wind.  This is in agreement with studies presented by \citet{2011A&amp;A...527A.119B} and \citet{2003ApJ...589..931N} who find CO emission generally consistent with stellar velocities, with average velocity difference $\Delta$v $\sim$ 3.5 km s$^{-1}$.  However, the possibility of a slow-moving disk wind, such as described in \citet{2011ApJ...733...84P}, cannot be fully ruled out as an explanation for the location of the absorbing CO gas in our targets.     
		
	The best-fit $^{12}$CO/$^{13}$CO ratio in our sample disks ranges from $\approx$~6~--~158 (See Table 4).  The errors in these ratios, however, are very large because of the order of magnitude errors in the column densities.  Our isotopic ratios are all consistent with the interstellar value of $^{12}$CO/$^{13}$CO $\sim$~70 \citep{2007ApJ...667.1002S} and the young stellar object environment value of $\sim$~100 found by \citet{2009ApJ...701..163S}, but the S/N and spectral resolution are not optimal for making a precise determination of the isotopic ratio in the disk.  
	
\subsection{CO Temperature and Density}	
	We now compare our CO absorption temperatures and column densities to other gas measurements of the inner disk in the literature.  The rotational excitation temperatures of our disks (300~--~700 K) agree well with the analyses of UV fluorescent CO emission lines ($460 \pm 250$ K) in \citet{2012ApJ...746...97S} for disks around similar CTTSs.  In contrast, the average rotational temperature of the full disk sample of \citet{2011ApJ...743..112S} is $\sim 1100$ K, albeit with a large range, which is higher than our sample average of $\sim 400$ K.  The difference is only marginally significant, although suggestive, due to the large uncertainties on the upper limit of the UV CO temperature distribution (see $\S 3.3$).  For a direct comparison, the model fit to the AA Tau near-IR CO emission by \citet{2011ApJ...743..112S} finds a rotation temperature of 950 K and log$_{10}$($N$(CO))$=18.6$, which are both higher than our parameter values for the same target.  \citet{2012ApJ...754...64H} find a CO column density for AA Tau of 1.2$\times$~10$^{19}$ cm$^{-2}$, also higher than our value.  We compare the \citet{2011ApJ...743..112S} and \citet{2012ApJ...754...64H} CO parameters for AA Tau to our fit by using their column density and temperature values in our model and plotting them with ours in Figure 6.  We assume a $^{12}$CO/$^{13}$CO ratio of 70, our $b$-value, and our velocity shift value for the other models.  The other models do not appear to be consistent with our models or data.  Because emission line studies probe a much larger region of the disk than the single line-of-sight of absorption line studies, it is not surprising that the \citet{2011ApJ...743..112S} study appears to probe a different molecular gas population.  The higher temperature gas may be located at a disk height larger than the line-of-sight sampled in our absorption line measurements.  The UV spectra may also be preferentially sensitive to lower column density gas due to extinction effects.  However, the apparent inconsistency between our model and the \citet{2012ApJ...754...64H} model for AA Tau is surprising because they are both absorption line studies.  It may be that the near-IR continuum emission and the UV continuum emission are produced in different locations, so that our lines-of-sight to them are different.  Alternatively, if the CO is located in the inner disk, then the difference may be related to the warp of the inner disk, which varies with the rotation of the star \citep{2007A&amp;A...463.1017B}.  The temperature fits to our data show that it is inconsistent with a temperature of 2500 K, the H$_{2}$ emission/absorption line temperature found by France et al. (2012a,b) in these disks, and the H$_{2}$ fluorescence temperature for TW Hya, modeled by \citet{2004ApJ...607..369H} and \citet{2005A&amp;A...438..923N}.  Assuming that both the CO and H$_{2}$ are in local thermodynamic equilibrium, the hot H$_{2}$ emission appears to be spatially separate from the CO absorption.  

\begin{figure} \figurenum{6}
\begin{center}
\plotone{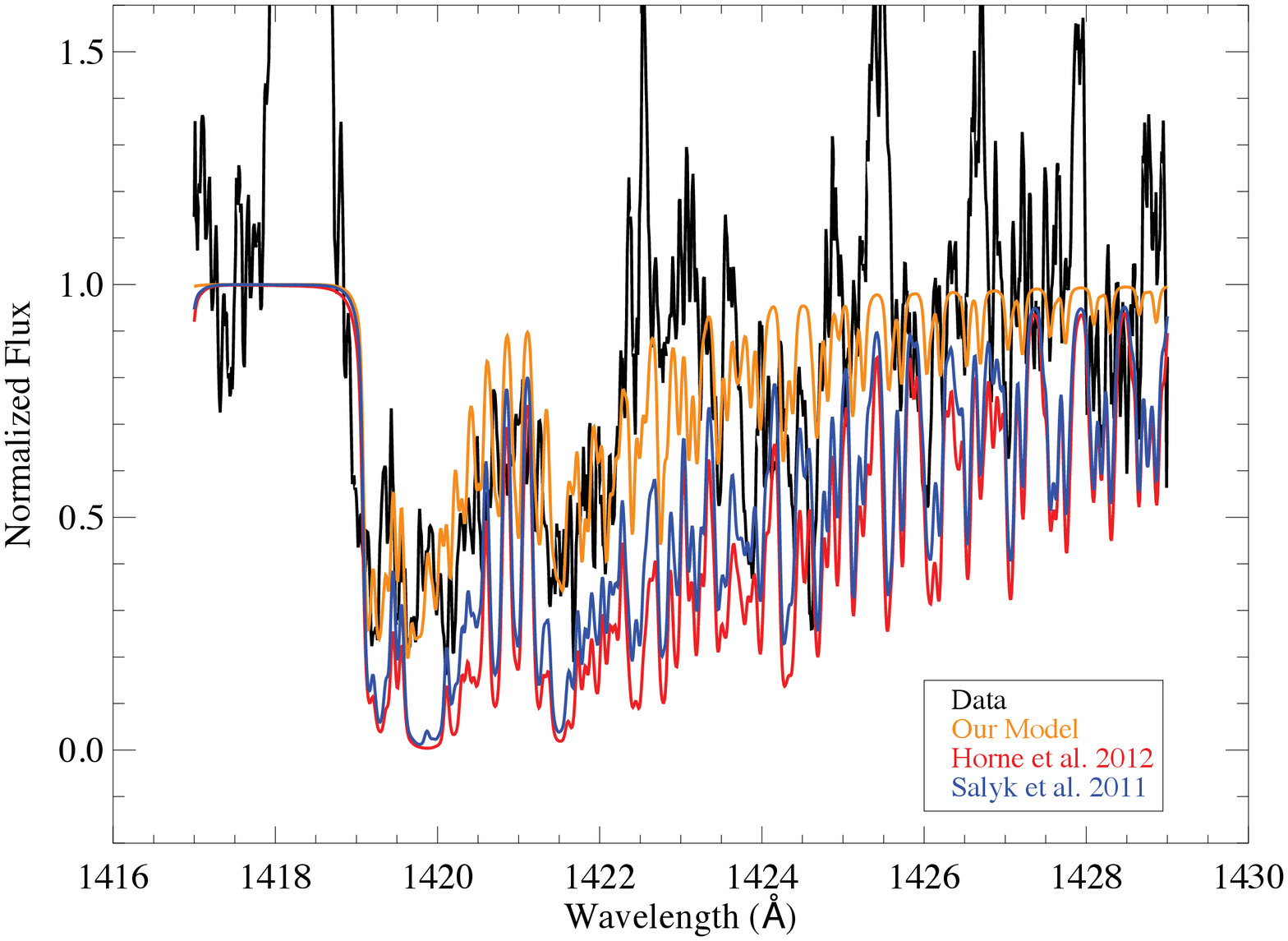}

\caption{
We use the temperature and column density values of \citet{2011ApJ...743..112S} and \citet{2012ApJ...754...64H} for AA Tau in our model and assume $^{12}$CO/$^{13}$CO $\sim$ 70 to compare their CO gas parameters to ours.  For the comparison, we use the (4 - 0) band of AA Tau.  The data is in black, our fit is in orange, the \citet{2011ApJ...743..112S} model is in blue, and the \citet{2012ApJ...754...64H} model is in red.  
}
\end{center}
\end{figure}
	      
	Assuming that our CO rotational temperatures are representative of gas kinetic temperatures T$_{gas}$ $>300$ K, we estimate the maximum radius where warm CO can be maintained at this temperature (from Figure C.2 of \citealt{2011A&amp;A...534A..44W}) to be $r_{CO} \approx 10$ AU.  This is consistent with the UV CO emission studied by \citet{2012ApJ...746...97S} who find that the emission arises from radii $\gtrsim$ 2 AU in similar disks.  Far-UV CO emission is spectrally unresolved from RECX-15, implying that $r_{CO} \geq 3$ AU.  We estimate the number density of the CO in these disks by dividing our column densities by this radius:  
\begin{equation}
  n_{CO} = \frac{N(^{12}CO)}{r_{CO}}
\end{equation}
The estimated CO number densities are in the range of $n_{CO} \sim 70 - 4000$ cm$^{-3}$, and are shown in Table 4.  With these densities, we computed the CO/H$_{2}$ ratio required for the CO to be thermalized up to $J$ = 25 (the highest $J$-state that could be reliably identified in our data).  If we assume that collisions with H$_{2}$ are the leading contributor to the CO level populations, and that the H$_{2}$ density is sufficiently high, then the kinetic temperature will equal the CO rotational temperature, $T_{rot}(CO)$.  Significant photoexcitiation would decouple the two temperatures.  We computed the H$_{2}$ critical density (the density where the spontaneous emission rate equals the collisional de-excitation rate) assuming an H$_{2}$ ortho/para ratio of 3 and using collision rates calculated by summing over all collisional routes \citep{2010ApJ...718.1062Y} downward out of level $J$ = 25.  The H$_{2}$ critical density for CO excitation in the range 300~--~750 K is $\approx$~(5.3~--~4.2)~$\times$~10$^{6}$ cm$^{-3}$.   The CO/H$_{2}$ ratios needed to thermalize the absorption lines are listed in Table 4.  They are upper limits and range from $1.6 \times 10^{-5}-7.9\times 10^{-4}$.  For the CO densities derived with the assumed Woitke maximum radius, typical interstellar translucent and dense cloud CO/H$_{2}$ ratios (10$^{-6}$~--~10$^{-4}$; \citealt{1994ApJ...428L..69L, 2007ApJ...658..446B}) are sufficient to maintain a thermal distribution for the observed CO absorption lines in five of the targets.  Although the derived CO densities for most of our targets suggest that the CO is thermalized, the sample is also consistent with sub-thermal excitation, within the uncertainties.  Sub-thermally populated lines would lead to the CO temperatures being underestimated.  If the absorbing CO population is in thermal equilibrium, then we conclude that the \emph{observed} CO/H$_{2}$ ratios derived by France et al. (2011a, 2012a) are not representative of the \emph{local} CO/H$_{2}$ ratios in the warm molecular disk surface.  	

\begin{deluxetable}{cccc}\tablenum{4}
\tabletypesize{\scriptsize}
\tablecaption{Protoplanetary Disk Warm Gas Parameters}
\tablewidth{0pt}
\tablehead{
\colhead{Object } & \colhead{$n_{CO}$} &  \colhead{CO/H$_{2}$\tablenotemark{a}} & \colhead{$^{12}$CO/$^{13}$CO} \\
 & cm$^{-3}$& & }

\startdata
AA Tau & $5.3^{+28.1}_{-4.5}\times 10^{2}$ & $< 1.1\times 10^{-4}$ & $25.1^{+373.0}_{-22.6}$\\
\\
DE Tau & $1.1^{+4.2}_{-0.9}\times 10^{3}$ & $< 2.3\times 10^{-4}$& $15.8^{+778.5}_{-14.2}$\\
\\
HN Tau & $6.6^{+26.7}_{-5.8}\times 10^{1}$ & $< 1.6\times 10^{-5}$ & $6.3^{+93.7}_{-5.0}$\\
\\
RECX-15 & $4.2^{+12.5}_{-3.8}\times 10^{3}$ & $< 7.9\times 10^{-4}$& $12.6^{+113.3}_{-11.8}$\\
\\
RW Aur & $5.3^{+21.2}_{-4.6}\times 10^{2}$ & $< 9.9\times 10^{-5}$& $158.5^{+635.8}_{-126.9}$\\
\\
SU Aur & $1.7^{+6.7}_{-1.5}\times 10^{3}$ & $< 3.1\times 10^{-4}$& $79.4^{+2432.5}_{-76.2}$\\
\enddata
\tablenotetext{a}{CO/H$_{2}$ ratio upper limit.  For CO/H$_{2}$ ratio less than this value, the absorption lines are thermalized up to $J$ = 25.}
\end{deluxetable}

	We compare our derived temperatures and approximated CO densities (which we change to a molecular hydrogen density with an assumed CO/H$_{2}$ ratio of $10^{-4}$) to the \citet{2011A&amp;A...534A..44W} model of RECX-15 as a check on the vertical disk structure.  Continuing our assumption that collisions with H$_{2}$ are the leading contributor to the CO level populations, the averaged hydrogen density will be dominated by H$_{2}$ and we can directly compare the plots in Figure C.2 of \citet{2011A&amp;A...534A..44W} with our temperatures and densities assuming the CO radius to be $r_{CO} \approx 10$ AU as above.  For each target's temperature and density pair, the height of the gas determined from the model are consistent with each other.  These heights range between $z/r$ $\sim 0.6$ to $\sim 0.7$ for the six targets, putting the gas in the flared upper disk atmosphere.            
	  
\subsection{Comparison of CO with System Parameters}

	Using the scale height model from $\S 4.2$, we calculate that only inclinations of $>79^{\circ}$ will intercept the A$_{V}=1$ surface of the disk, which is larger than the inclinations in Table 1.  At the inclinations of the targets, the sightlines are not probing into the depth of the disk where the visual extinction exceeds unity, which indicates that our CO is likely well above the A$_{V}=1$ disk surface.  However, using a dimensionless height of $z/r$ = $0.6$, the disk would be intercepted for inclinations $\gtrsim 60^{\circ}$.  This increases our estimate on the unconstrained HN Tau inclination.  The lower limit on inclinations of $\gtrsim 60^{\circ}$ is compatible with most of our disk inclinations, with the notable exception of DE Tau.  The inclination of DE Tau is too small for us to be observing the gas in absorption, yet we clearly see the absorption bands in the data.  DE Tau may be a candidate for a protoplanetary disk with a slow-moving molecular disk wind with a more face-on inclination and blue-shifted lines \citep{2011ApJ...733...84P}.  DF Tau, another target in the \citet{2012ApJ...756..171F} sample known to have CO \citep{2003ApJ...589..931N, 2011ApJ...743..112S, 2012ApJ...746...97S} and thought to have a high inclination ($85^{\circ}$; \citealt{2001ApJ...561.1060J}), does not show UV CO absorption, which is surprising.  DF Tau has an anomalously large radius for a $10^{6}$ Myr star (3.37 $R_{\sun}$, \citealt{1998ApJ...492..323G}), which leads to the large inclination value.  The spectrum for DF Tau is shown in Figure 7 with the fit for AA Tau overplotted to illustrate the non-detection.  A revised determination of the inclination of DE Tau and DF Tau, and possibly the other sources in the sample, would be useful.  With notable exceptions, for our mid- to high-inclination disks known to contain CO, we see absorption if the gas is located at a height of $z/r$ $\gtrsim 0.6$, which is compatible with our derived temperatures and estimated densities.  A cartoon of this geometry is shown in Figure 8.   

\begin{figure} \figurenum{7}
\begin{center}
\plotone{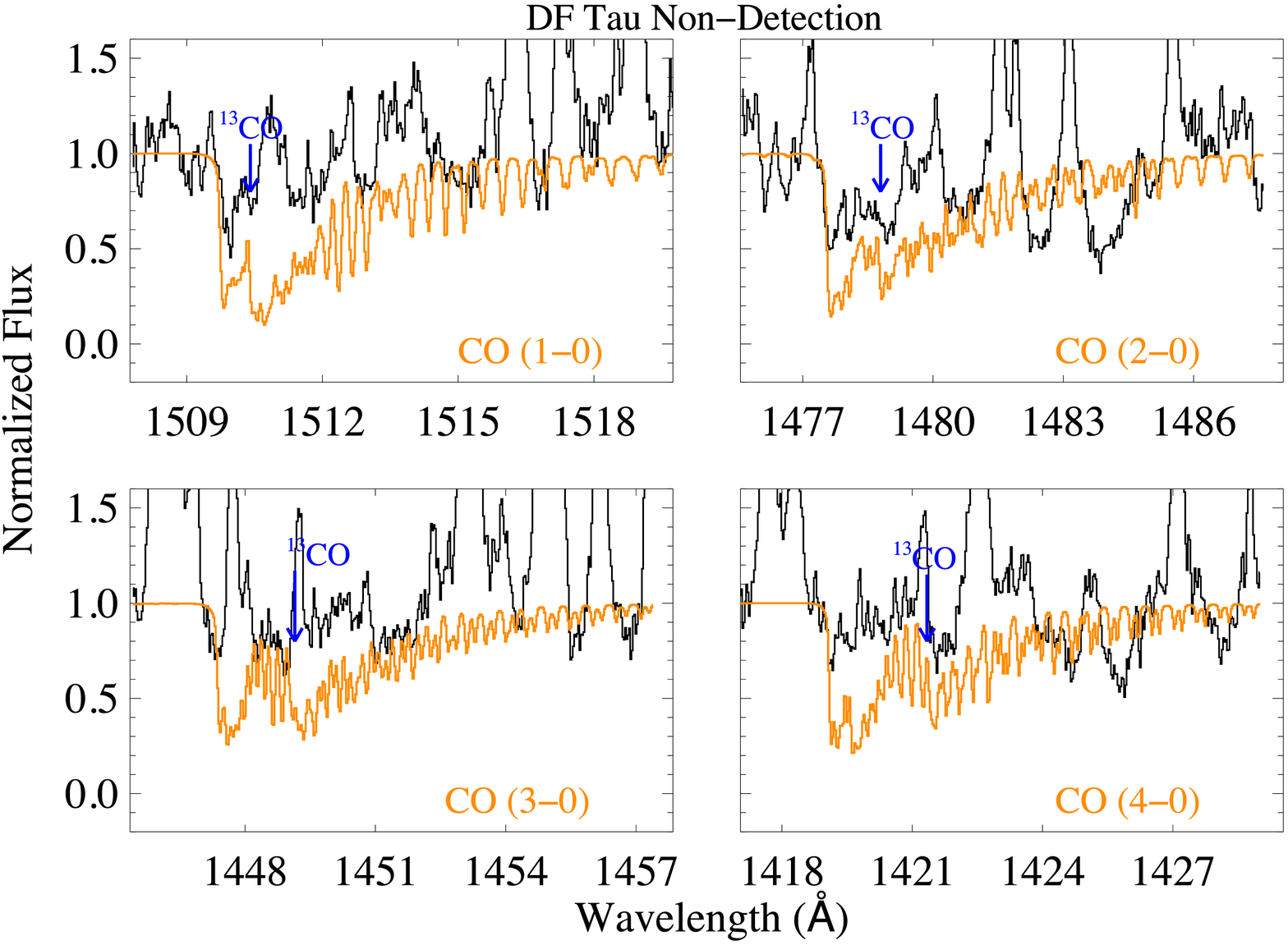}

\caption{
COS data for DF Tau with the model fit of AA Tau overplotted, showing the absence of CO absorption in the DF Tau spectrum (DF Tau data in black and best-fit model for AA Tau in orange).  The $^{13}$CO bandheads are marked to show the two different CO species.  
}
\end{center}
\end{figure}

\begin{figure} \figurenum{8}
\begin{center}
\plotone{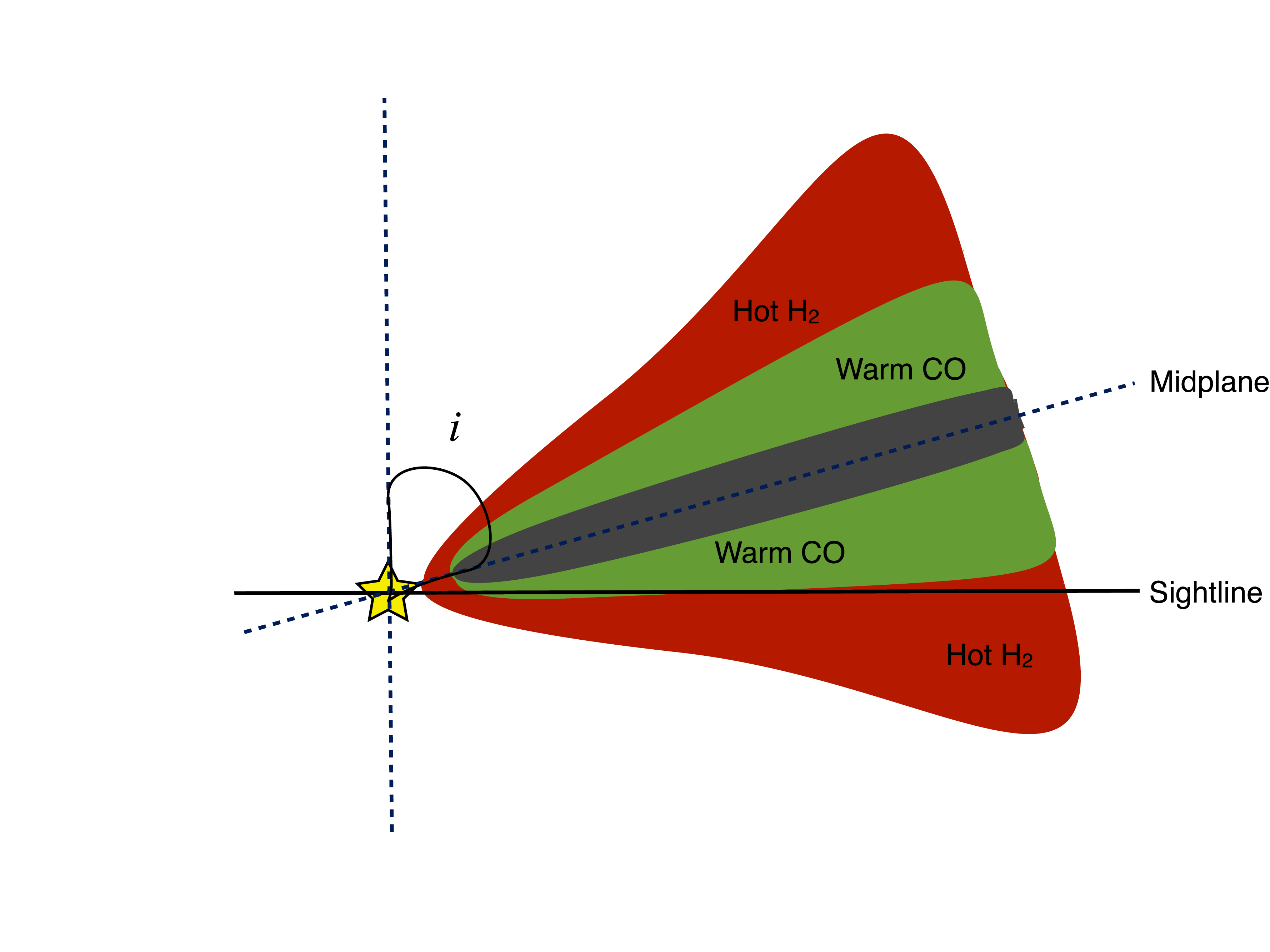}

\caption{
A schematic representation of the line-of-sight geometry for the inner region of T Tauri star disks.  The red layer is observed as low-density, hot (T~$\sim$~2500 K) H$_{2}$ emission and absorption \citep{2004ApJ...607..369H, 2011ApJ...730L..10Y, 2012ApJ...744...22F}; the green material is observed as intermediate density, warm (T~$\sim$~500 K) CO emission and absorption; and the dark midplane is inaccessible to UV observations.   
}
\end{center}
\end{figure}
	
	In Figure 9, we compare the column density and rotational temperature of CO with the mass accretion rate and disk inclination.  We do not see any strong correlations between these parameters, though there may be a slight decrease in temperature with increasing disk inclination.  At higher inclinations, we are observing the disk more edge-on so that the line-of-sight passes through the colder, denser material lower in the disk atmosphere.   

\begin{figure} \figurenum{9}
\begin{center}
\plotone{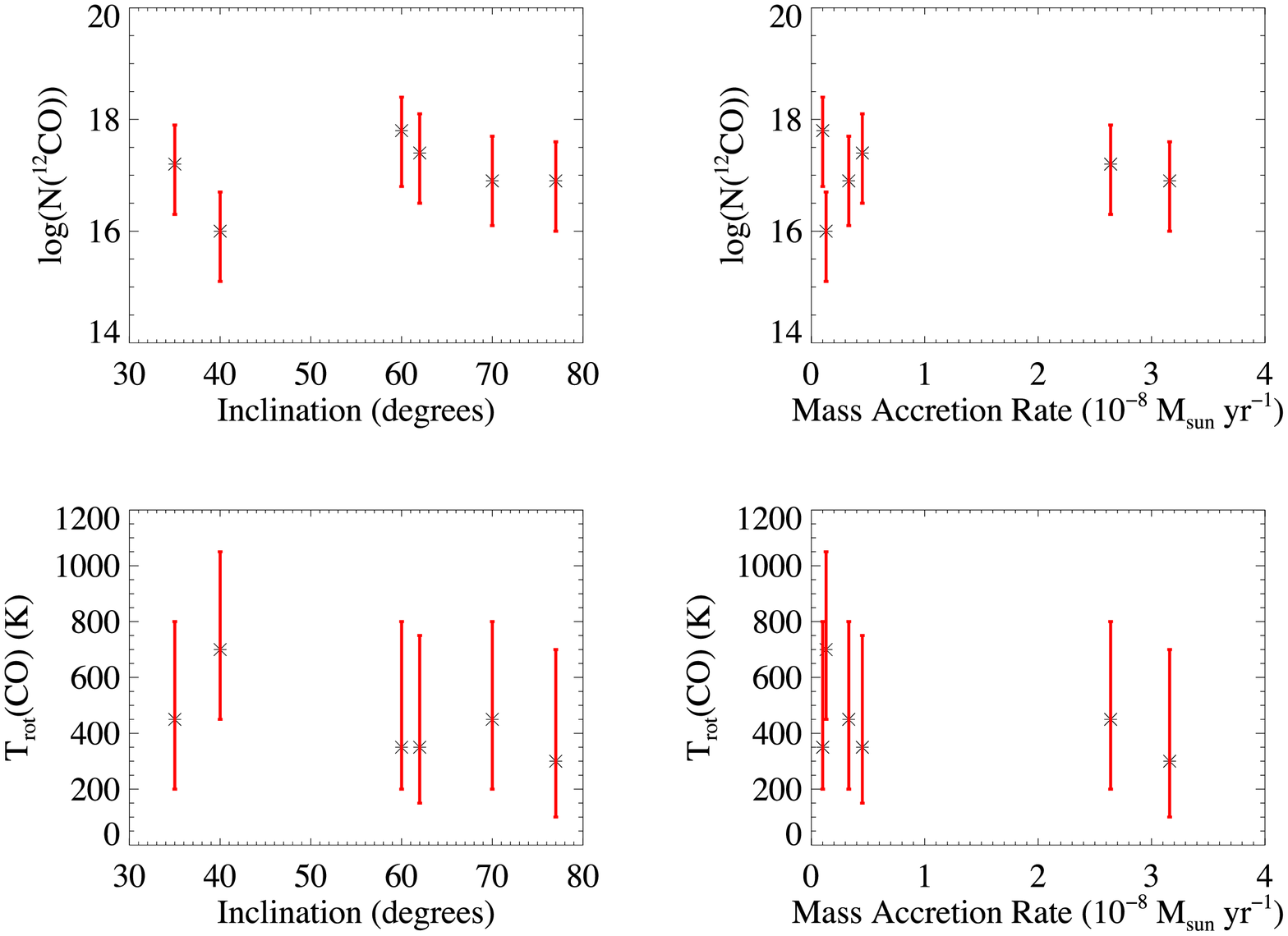}

\caption{
Correlations of the target parameters and best-fit parameters.  
}
\end{center}
\end{figure}
		
\section{Conclusions}	
	We present model fits to ultraviolet CO absorption lines in six protoplanetary systems.  We find results in AA Tau and HN Tau that are consistent with previous work and extend our analysis to four new targets obtained with $HST$-COS far-UV G130M and G160M modes.  We find CO rotational temperatures in the range 300 - 700 K, which agree well with UV CO fluorescence rotational temperatures.  However, our temperatures are cooler than inner disk CO temperatures obtained from the analyses of IR CO emission spectra and warmer than the \citet{2006ApJ...646..342R} CO temperatures from IR CO absorption.  The IR emission data may be probing a different gas population at smaller disk radii, while the IR absorption data is likely probing gas at larger radii.  The measured velocities of the CO absorbing gas rule out an origin in a fast-moving disk wind.  Our derived temperatures and approximated densities of the gas are consistent with models for disk heights of $z/r$ $\sim 0.6$.  This CO location constrains the inclination of our disks to be $\gtrsim 60^{\circ}$ in order to intercept the absorbing gas in the model. 
	
	We note that the present analysis is roughly at the limit of what is feasible with the current generation of $HST$ instrumentation.  Higher spectral resolution would improve our molecular parameter determination significantly.  Unfortunately, observations at these flux levels with $HST$-STIS E140M mode are not feasible.  A new observational capability will be required to derive more robust disk parameters from CO absorption line observations of CTTSs.  However, future observations with the new COS G130M $\lambda$1222 mode should be able to directly probe warm H$_{2}$ absorption in these disks at wavelengths around 1100~\AA. These observations would allow us to directly determine the value of the CO/H$_{2}$ ratio in moderate-to-high inclination protoplanetary disks.    	        
	                 
\acknowledgments
	We thank M. Eidelsberg for providing the CO oscillator strengths and rovibrational line wavelengths used in this work.  This work made use of data from HST guest observing program 11616 and was supported by NASA grants NNX08AC146 and NAS5-98043 to the University of Colorado at Boulder.


  \bibliography{ms}

\end{document}